\documentclass[12pt,a4paper]{article}
\usepackage{amssymb}
\usepackage{amsfonts}
\usepackage{amsmath}

\input{tcilatex}
\begin{document}

\title{Does the Superluminal Neutrino Uncover Torsion?}
\author{M.B. Altaie \\
Department of Physics, Yarmouk University, 21163 Irbid, Jordan}
\maketitle

\begin{abstract}
I investigate the possibility of the propagation of neutrino with
superluminal speed through matter in the context of the relation between
gravity, spin and tortion. Using a lemma of Penrose and earlier works on the
relation between spin, torsion and gravity I glimpse on a frame work in
which superluminal speed of the neutrinos moving through matter become
possible. In presence of torsoin neutrinos are found to follow spacelike
geodesics by tunneling through the light cone into the spacelike region and
consequently appear to have superluminal speed in our timelike world,
whereas photons always follow null geodesics. This frame work may set new
frontiers for spacetime physics.
\end{abstract}

\section{Introduction}

Measuring the flight time of the muon neutrinos, with average energy of $%
17.5 $ GeV, the OPERA collaboration reported that they have measured a
superluminal speed for these neutrinos \cite{Adam}. They claims that their
experiment gives a travel time for the ultrarelativistic neutrinos which is
about $60$ ns less than expected when compared to the speed of light. This
means that neutrinos propagate with superluminal speed with $\delta
v=(v-c)/c\sim 2.5\times 10^{-5}$ where $c$ is the speed of light in vacuum.
Earlier experiments on high energy neutrinos like the MINOS project \cite{MS}
have shown that $\delta v\sim 5.1\times 10^{-5}$. Astronomical detection of
neutrinos from the supernova SN1987a has given much less figure of about $%
\delta v\sim 2\times 10^{-9}$.

Several interpretations for the OPERA claim has be published in pre-prints
recently, most of them are \textit{ad hoc} suggestions that lacks rigor and
firm foundations. In fact, a more profound explanation is needed to explain
the Opera, MINOS and the SN1987a results, since if true this result will
have immense implications on the understanding of spacetime and particles,
including the question of general covariance. In this letter I am not going
to present any full fledge theory on the subject but will hint on some views
that might enable tackle the subject from another point of view toward an
approach that might explain these results.

\section{Torsion, Spin and Gravity}

Symmetry is one of the most beautiful aspects of nature that our minds may
envision. However, in a deeper scientific prospect symmetry stands as the
unifying pedestal on which laws of nature rests; behind every symmetry there
hides a conservation law that keeps the balance. Forces and potentials may
break the symmetry but then other forces and potentials come into play to
restore symmetry in a wider scope and keep the balance in nature.

Several alternatives to the theory of general relativity has been suggested
since it was proposed in 1915. Perhaps the most interesting of these was
Cartan's introduction of torsion as the antisymmetric part of an asymmetric
affine connections \cite{Cartan}. Cartan recognized the tensor character of
torsion and developed a differential geometric formulation and he had some
ideas about torsion of the spacetime being connected with the intrinsic
angular momentum of matter and later Schr\"{o}dinger tried to develop a
unified theory of gravity and electromagnetism in 1943 where torsion was
related to electromagnetic potential \cite{Sch}, consequently Schr\"{o}%
dinger found that photons acquire a non-zero rest mass (for an overview of
the literature see \cite{Hh} and for more recent review see \cite{Hmm}).
This culminated later into the formulation of what is called the $U_{4}$
theory by Kibble \cite{Kib} and \cite{Sc} which came in the context of the
gauge approach to gravity. Both Kibble and Sciama arrived at a set of field
equations and laid down the basic structure of $U_{4}$ theory. Further
development of this approach was taken by Hehl, Von der Heyde and Kerlick 
\cite{Hh2}, and Trautman \cite{Traut} and others. The birth of local gauge
theory in the 1950s breathed new life into torsion. With the first attempts
by Utiyama \cite{Ut} paving some ground, and Sciama \cite{Sc} emphasizing
torsion as being related to spin. Kibble \cite{Kib} showed how to describe
gravity with torsion as a local gauge theory of the Poincare' group, and by
1976 Hehl \textit{et al} \cite{Hh} formulated a gravitation theory with
torsion as resulting from local Poincare' gauge invariance. The beauty and
success of global Lorentz invariance was generalized with modern gauge
principles to form a compelling new picture. The two Casimir invariants of
the Poincare' group, the square of the translation operator $P^{2}$ and
Pauli--Lubanski spin operator $L^{2}$ found perfect interpretations in a
theory of gravity with torsion: generalizing the notion that mass curves
space, now we also have spin giving rise to torsion. It was anticipated that
having formulated gravitation as a gauge theory the charm and success of the
quantization of the $SU(n)$ theories might rub off on gravity, yielding a
quantizable theory. However, it was disappointing that the early predictions
indicated that torsion forces were too weak to measure, and, in some
formulations, torsion did not even propagate into vacuum. However if the
torsion tensor is to be taken as being a gradient of a scalar potential then
it could propagate in vacuum as shown by Hammond \cite{Hmm}.

\subsection{Gravity with Torsion}

Gravity, according to Einstein, is a curvature of the spacetime, and matter
would tell spacetime how to curve. This was the view according to Wheeler.
However, matter seem to have other effects on spacetime other than
curvature. If we would understand curvature in terms of the length of
trajectories and durations of time the we can only do that in a comparative
context. But if we have to understand curvature in its intrinsic character
then we have to look for an intrinsic context within which we can appreciate
the meaning of curvature. This is usually done through what we call
transporting a vector parallel to itself a long a closed trajectory or a
surface in the spacetime. If the vector is preserved throughout the
displacement precisely then the curvature of the spacetime is zero and the
spacetime is called \emph{flat}. Otherwise, if there would be a difference
in the magnitude or the direction of the vector throughout the trip then in
this case we say that the spacetime is curved. In order to preserve the
symmetry in curved space a covariant derivative should replace the ordinary
derivative to express the infinitesimal translation. Curvature is the
property of the spacetime endowed with gravity, such a spacetime can be
available outside the matter source. However, once matter (in the form of a
test particle for example) is introduced in this spacetime new effects may
appear and among these is torsion. The spacetime get rapped in a chiral
manner causing light cones to shrink in presence of torsion. This will
change the teleparallel behavior of the spacetime, and accordingly the
metric and the covariant derivative all will change.

The behavior of sticks and clocks in torsion free curved spacetime is well
developed in general relativity which considers a Riemannian spacetime with
all the symmetries enjoyed within like metricity, covariance and
conservations of energy and momentum. In such a spacetime the covariant
derivative 
\begin{equation}
\nabla _{\mu }=\partial _{\mu }+\Gamma _{\mu \nu }^{\sigma }  \label{q8}
\end{equation}%
is introduced in order to account for the locality of the gravitational
potentials and to safeguard the general covariance of the translational
symmetry. The affine connections $\Gamma _{\mu \nu }^{\sigma }$ are assumed
to be symmetric in $\mu $ and $\nu $ and the metric tensor is taken to be
divergenless expressing the conservation of spacetime. Such assumptions have
constrained the spacetime to be torsion free. Now, if we have to take care
of the chiral symmetry of the spacetime and look for the introduction of
spinning matter we have to introduce torsion; for torsion is the object that
is related in essence to chirality. For this gaol we should expect the
Riemannian affine connections to be modified. Indeed these are given by 
\begin{equation}
\hat{\Gamma}_{\mu \nu }^{\sigma }=\Gamma _{\mu \nu }^{\sigma }-K_{\mu \nu
}^{..\sigma }  \label{q9}
\end{equation}%
where $\Gamma _{\mu \nu }^{\sigma }$ are the usual Riemann-Christoffel
symbols and $K_{\mu \nu }^{..\sigma }$ is the cotorsion tensor related to
the torsion by 
\begin{equation}
S_{\mu \nu }^{..\sigma }=\Gamma _{\lbrack \mu \nu ]}^{\sigma }=-K_{[\mu \nu
]}^{..\sigma }  \label{q10}
\end{equation}

The Covariant derivative in the Riemann-Cartan space takes the form

\begin{equation}
\hat{\nabla}_{\mu }=\partial _{\mu }+\hat{\Gamma}_{\mu \nu }^{\sigma }
\label{q11}
\end{equation}

The fact that torsion is basically antisymmetric motivates one to foresee
some connection with the rotational properties of the spacetime, and perhaps
this what Cartan had in mind originally. This what would introduce the
Poincare group into the picture (for more details see sec.IV of Ref \cite{Hh}%
).

From the definition of the covariant derivative in (\ref{q9}) and the
properties of $K_{\mu \nu }^{..\sigma }$ it is easy to see that $\hat{\nabla}%
_{\mu }g_{\rho \sigma }=0$, a property by which the metricity of the
Riemann-Cartan spacetime is maintained.

\subsection{The variational consideration}

The total action of gravity with matter and torsion is taken to be extremal
according to 
\begin{equation}
\delta (I_{G}+I_{m}+I_{s})=0  \label{q1}
\end{equation}%
where $I_{G}$ is the geometrical action given by 
\begin{equation}
I_{G}=\frac{c^{4}}{16\pi G}\int \sqrt{-g}Rd^{4}x  \label{q2}
\end{equation}%
where $R$ is the scalar curvature $g$ is the determinant of the metric
tensor. $I_{m}$ is the matter action given by 
\begin{equation}
I_{m}=\frac{1}{2}\int \sqrt{-g}T^{\mu \nu }\delta g_{\mu \nu }dx^{4}
\label{q3}
\end{equation}%
with $T^{\mu \nu }$ being the energy-momentum tensor. $I_{s}$ is the torsion
action given by 
\begin{equation}
I_{s}=\frac{1}{2}\int \sqrt{-g}\mu _{\sigma }^{\mu \nu }\delta S_{\mu \nu
}^{\sigma }dx^{4}  \label{q4}
\end{equation}%
where $\mu _{\sigma }^{\mu \nu }$ is the spin-energy potential of matter.
From (\ref{q1}) the field equations can be obtained as 
\begin{equation}
G^{\mu \nu }-\hat{\nabla}_{\beta }\left( T^{\mu \nu \beta }+T^{\beta \mu \nu
}+T^{\beta \nu \mu }\right) =\frac{16\pi G}{c^{4}}T^{\mu \nu }  \label{q5}
\end{equation}%
where 
\begin{equation}
T^{\alpha \beta \sigma }=S^{\alpha \beta \sigma }+S^{\beta }g^{\alpha \sigma
}-S^{\beta }g^{\beta \sigma }  \label{q6}
\end{equation}%
is the modified torsion tensor, $S^{\alpha \beta \sigma }$ is the torsion
tensor and $S^{\beta }=S_{\sigma }^{\beta \sigma }$ is the torsion trace or
the torsion vector. The torsion tensor is related to matter potentials by
the equations%
\begin{equation}
S_{\alpha \beta \gamma }=\frac{16\pi G}{c^{4}}\left( \mu _{\gamma \lbrack
\alpha \beta ]}+\mu _{\lbrack \alpha }g_{\beta ]\gamma }\right)   \label{q7}
\end{equation}%
where $\mu _{\alpha }=\mu _{\alpha \sigma }^{\sigma }$. Clearly torsion
vanishes in vacuum, therefore exterior to matter source the field equations (%
\ref{q5}) reduces to the standard Einstein field equations and their
solutions will produce the same geodesics that are produced in Riemannian
space.

\section{Propagation of Matter Fields in U$_{4}$}

The propagation of matter fields in the background of Riemann-Cartan
spacetime has been studied by few authors and its is shown that scalar
field, which has no spin, neither feel nor produce torsion. Photons in $U_{4}
$ are unaffected by the presence of torsion and consequently the causal
structure of a $U_{4}$ spacetime is determined completely by the conformal
metric structure of the spacetime \cite{Hh}. The propagation of massive
Dirac field in $U_{4}$ was studied by Jurgen \cite{Jurgen} and it was shown
that particles follow non-geodesic trajectories. The force equation is given
as%
\begin{equation}
mv_{{}}^{\nu }\hat{\nabla}_{\nu }v_{\mu }=\frac{1}{2}\left( \frac{\hslash }{2%
}\right) \overset{}{\hat{R}_{\mu \sigma \rho \lambda }}\bar{b}_{0}\sigma
^{\rho \lambda }b_{0}u^{\sigma }  \label{q12}
\end{equation}%
where $\sigma ^{\rho \lambda }=i[\gamma ^{\rho },\gamma ^{\lambda }]$, $b_{0}
$ is a spinor column matrix, and $\bar{b}_{0}$ is its conjugate. Clearly the
force in Eqn. (\ref{q12}) is taken to the order of $\hslash $, if this order
is dropped then the trajectories of Dirac particles will follow geodesics in
the Riemann-Cartan spacetime. However, this force may also hint a
verification of torsion through a suitable experiments.

If Neutrinos are to fly with superluminal speed then they have to follow a
spacelike trajectories \cite{Wein}, and according to Penrose such
trajectories are possible if torsion exist \cite{Pen}. In this case, as
remarked by Penrose, the light cone will be a timelike surface with respect
to $\hat{\nabla}$, but would curl into the inside of the light cone with
respect to $\nabla .$Therefore, the trajectory of the neutrino will escape
from inside to outside the light cone. The reason why the observed velocity
of the neutrinos gained marginal velocity above c is that their torsion
potential is low.

The supernova SN1987a neutrinos did not show any significant superluminal
speed because it was propagated mostly in vacuum. This is where torsion
would not play any significant role. In order to testify our suggestion
presented here and other suggestions other experiments with longer
base-lines are needed in order to check the path dependence of this effect.

The situation suggested here is geometrically similar to that which is
suggested by the bimetric relativity. Recently Moffat \cite{Mof} suggested
that a bimetric structure of the spacetime might offer an explanation for
the superluminal neutrino. In fact it is clear that the underlaying gauge $%
\psi _{\mu }$ variance is playing the underlaying connecting substratum in
both approaches. This might be shown in more details if we consider the
calculation of $\hat{\Gamma}_{\mu \nu }^{\sigma }$ from the given bimetrical
form 
\begin{equation*}
\hat{g}_{\mu \nu }=g_{\mu \nu }+\beta \psi _{\mu }\psi _{\nu }
\end{equation*}%
and calculate $\hat{\Gamma}_{\mu \nu }^{\sigma }$ where we find a similar
relation to (\ref{q9}). In this case the torsion tensor will get its
definition in terms of the bivector $\psi _{\mu }\psi _{\nu }$. This analogy
will help investigate the qualitative and quantitative differences in the
physical effects that both theories may predict.

The issue of how neutrinos or other particles can cross the light barrier
into the spacelike region is something that has to be investigated more
profoundly. However, such tunneling into the spacelike region will no doubt
assert the massiveness of the neutrino no matter how small is its mass,
otherwise . The very recent suggestion of Ahluwalia, Horvath and Schritt 
\cite{Ah} for the mass of the neutrino spacies in the context of discussing
a possible limit on the neutrino masses based on the measured speed of the
neutrino is consistent with this view.

The point which concerns theoretical physicists is that crossing the light
barrier for a superluminal speed would break Lorentz covariance, and in fact
this will not be a problem. We are accustomed to view physics in the
timelike regions of the spacetime. This is motivated by the wish to maintain
causality in the relativistic meaning. But if we take into consideration all
the spacetime with its timelike and spacelike regions to be under a more
general transformation law, then the physics might get easier and better
understood. Indeed timelike and spacelike regions can be interchanged using
the duality transformation in order to get the full picture of the physical
phenomena. For example if we consider the duality transformation of the
electromagnetic field we find that on a special choice of the $\theta $ the
electric monopole is replaced by the magnetic monopole and vise versa. If
the duality transformations are generalized to cover spacetime coordinates
then it would imply interchanging spacelike and timelike regions.
Consequently, I would suggest here that magnetic monopoles does exist, but
only in the spacelike region of the spacetime. Beside this and since
magnetic monopoles are necessary for the quantization of charge as shown by
Dirac \cite{Dirac} then it would be of interest to understand the action of
these monopoles from under the carpet of our timelike world instead of
looking for them in labs. In this context comes the neccessity to complement
the timelike regions of the light cone with the spacelike regions to form a
complete manifold for physics. Within this scope the role of exotic objects
like magnetic monopoles, tachyons and supersymmetric particles may be better
understood, and the coupling of mass to curvature, electromagnetic field to
null geodesics, torsion and spin to gravity get clearer to complete the
picture of the world. \ \ 

\section{Conclusions}

Spinning neutrinos couples to torsion in spacetime. Torsion is a spatial
property that is reflected in the translational structure of the spacetime.
If superluminal speed of neutrinos proves to be true, then one might
conjecture that they follow spacelike geodesics and consequently their
superluminal speed is understood. In this context it might be quite possible
to use neutrino-experiment as an effective tool for probing the properties
of spacetime, specifically this can be used to investigate the torsion
effects (if any) at different energies of the probing particles. A long
base-line experiments are needed now for two reasons: (i) to make sure that
neutrinos are propagated with a speed faster than light and (ii) to see if
that effect is distance dependent. Such investigations will enable physics
get new frontiers that might help bring gravity and quantum mechanics closer
to meet on the bridge of the intrinsic spin of elementary particles.


\begin{thebibliography}{99}
\bibitem{Adam} T. Adam \textit{et al}., arXiv:1109.4897 [hep-exp].

\bibitem{MS} P. Adamson \textit{et al}., Phys.Rev. D \textbf{76} 072005,
(2007).

\bibitem{Cartan} Cartan E. C. R. Acad. Sci., Paris \textbf{174}, 593, (1922).

\bibitem{Sch} E. Schrodinger, , Proc. R. Ir. Acad. A \textbf{49}
43--58,(1943).

\bibitem{Hh} F. W. Hehl, P. von der Heyde , G. D. Kerlick and J. M. Nester ,
Rev. Mod. Phys. \textbf{48} 393--416, (1976).

\bibitem{Hmm} R. Hammond, Rep. Prog. Phys. \textbf{65}, 599-649, (2002).

\bibitem{Ut} R. Utiyama, Phys. Rev. \textbf{101} 1597--607, (1956) .

\bibitem{Kib} T. W B Kibble J. Math. Phys. \textbf{2,} 212--21,(1961).

\bibitem{Sc} D. W. Sciama Rev. Mod. Phys. \textbf{36} 463--9, (1964).

\bibitem{Hh2} Hehl F. W., von der Heyde P., Kerlick G. D., Phys. Rev D%
\textbf{10}, 1066. (1974).

\bibitem{Traut} A. Trautman, Ann. N. Y. Acad. Sci. \textbf{262}, 241, (1975).

\bibitem{Jurgen} J. Audretsch, Phys. Rev. D \textbf{42}, 1470-77, (1981).

\bibitem{Wein} S.Weinberg, Gravitation and Cosmology: Principles and
Applications of the General Theory of Relativity. John Wiley \& Sons,
NewYork, USA,1972.

\bibitem{Pen} R. Penrose, Techniques of Differential Topology in Relativity,
Society for Industrial and Applied Mathematics, Philadelphia, (1972).

\bibitem{Mof} Moffat, arXiv: 1110.1330v3 [hep-ph].

\bibitem{Ah} D. V. Ahluwalia, S. P. Hovrvath and D. Schritt, arXive:
1110.1162v1 [hep-ph].

\bibitem{Dirac} P. A. M. Dirac, Phys. Rev. \textbf{74}, 817-830, (1948).
\end{thebibliography}
\end{document}